\documentstyle[aaspp4]{article}

\slugcomment{Submitted to The Astrophysical Journal}

\lefthead{Borione, {\em et al.}}
\righthead{Gamma rays from the Galactic Plane at 300 TeV}

\begin{document}

\title{\hspace*{5in}{\tt \small EFI~Preprint~96-43}\\
\ \\
Constraints on Gamma-ray Emission from\\ 
the Galactic Plane at 300 TeV}
\author{
A. Borione,\altaffilmark{1}
M.A. Catanese,\altaffilmark{2,4}
M.C. Chantell,\altaffilmark{1}
C.E. Covault,\altaffilmark{1}
J.W. Cronin,\altaffilmark{1}
B.E. Fick,\altaffilmark{1}
L.F. Fortson,\altaffilmark{1}
J. Fowler,\altaffilmark{1}
M.A.K. Glasmacher,\altaffilmark{2}
K.D. Green,\altaffilmark{1}
D.B. Kieda,\altaffilmark{3}
J. Matthews,\altaffilmark{2}
B.J. Newport,\altaffilmark{1}
D. Nitz,\altaffilmark{2}
R.A. Ong,\altaffilmark{1}
S. Oser,\altaffilmark{1}
D. Sinclair,\altaffilmark{2}
J.C. van der Velde\altaffilmark{2}
}

\altaffiltext{1}{Enrico Fermi Institute, University of Chicago, Chicago, IL 60637}

\altaffiltext{2}{Department of Physics, University of Michigan, Ann Arbor, MI 48109}
\altaffiltext{3}{Department of Physics, University of Utah, Salt Lake City, UT 84112}
\altaffiltext{4}{Present Address: Department of Physics and 
Astronomy, Iowa State University, Ames, IA 50011}

\authoremail{covault@hep.uchicago.edu}
\begin{abstract}

We describe a new search for diffuse ultrahigh energy gamma-ray
emission associated with molecular clouds in the galactic disk. The
Chicago Air Shower Array (CASA), operating in coincidence with the
Michigan muon array (MIA), has recorded over $2.2\times10^{9}$ air showers
from April 4, 1990 to October 7, 1995.  We search for gamma rays based
upon the muon content of air showers arriving from the direction of
the galactic plane.  We find no significant evidence for diffuse
gamma-ray emission, and we set an upper limit on the ratio of
gamma rays to normal hadronic cosmic rays at less than 2.4$\times
10^{-5}$ at 310 TeV (90\% confidence limit) from the galactic plane
region: $(50\arcdeg <\ell_{\mbox{\scriptsize II}} < 200\arcdeg;
-5\arcdeg < b_{\mbox{\scriptsize II}} < 5\arcdeg)$.  This limit places
a strong constraint on models for emission from molecular clouds in
the galaxy.  We rule out significant spectral hardening in the
outer galaxy, and conclude that emission from the plane at these
energies is likely to be dominated by the decay of neutral pions
resulting from cosmic rays interactions with passive target gas
molecules.

\end{abstract}


\sloppy

\section{Introduction}

Diffuse gamma rays arriving from regions of enhanced density, such as
giant molecular clouds, are powerful tracers of the interactions of
cosmic rays with matter in the galaxy.  We expect diffuse emission
from molecular cloud regions where the permeating flux of high energy
cosmic rays interacts with the nuclei of gas atoms (mostly hydrogen)
to produce new hadronic particles, including neutral pions, which
subsequently decay into gamma rays.  In this manner, the gas in
molecular clouds acts as a passive target, converting some fraction of
impinging cosmic rays into gamma rays.  Such diffuse emission has been
recorded by several space-borne detectors including the EGRET
experiment aboard NASA's Compton Gamma Ray Observatory -- indeed the
all-sky flux at 100 MeV is largely ($\sim 90$\%) diffuse
emission from molecular clouds in the galactic plane.  Most of this
emission is contained within a narrow band along the galactic equator
($b_{\mbox{\scriptsize II}} < \pm 5\arcdeg$), and is sharply
concentrated towards the galactic center.  The EGRET results have been
used, together with radio data, to develop a three-dimensional model
of both the gas and cosmic ray densities in the galaxy (\cite{Ber93},
\cite{Hun97}).  This model, which estimates the contributions to
diffuse emission from electron bremsstrahlung, nucleon-nucleon
interactions, and inverse Compton processes, accurately matches the
observed emission in detail as seen by EGRET for all galactic
longitudes over the energy range from 30 MeV to about 1 GeV. (The
model deviates somewhat from observations at energies above 1 GeV, possibly due
to a change in the cosmic ray spectrum; see \cite{Mori97}).  The
EGRET results support a galactic origin of cosmic rays,
presumably as a result of shock acceleration in supernova remnants
(SNR).

Much less is known about the nature and distribution of cosmic rays at
energies above 100 TeV ($10^{14}$ eV).  Past efforts have mostly 
focused on searches for point sources of gamma rays using large-area
ground-based air shower detectors.  However, after many sensitive
searches, there exists no compelling evidence for the detection of
even a single persistent gamma-ray point source at these energies (see
{\em e.g.}, \cite{Alex93}, \cite{Cro93}, \cite{casa_binary}, \cite{casa_crab}).

We can estimate the ultrahigh energy diffuse flux expected from
clouds in the galaxy by approximating the cosmic ray density and the
matter density within any particular molecular cloud as uniform.  In
this case, the diffuse gamma-ray flux from this interaction can be
expressed as:
\begin{equation} J_{\gamma}(> E) = \frac{1}{4\pi}q_{\gamma}(>E)
\cdot <N_{H}>,\end{equation} 
where $<N_{H}>$ is the total hydrogen column density in a given
direction through the cloud and $q_{\gamma}(>E)$ is the source
function for gamma-ray emissivity:
\begin{equation}
q_{\gamma}(>E)= 4\pi J_{\mbox{\tiny CR}}(> E) \cdot 
{\cal F}(\sigma_{\mbox{\footnotesize inel}}, \gamma, f_A).
\end{equation}
Here $J_{\mbox{\tiny CR}}(> E)$ is the ambient integral flux of normal
hadronic cosmic rays and ${\cal F}$ is a function of the total
inelastic cross section, $\sigma_{\mbox{\footnotesize inel}}$, the
integral spectral index of cosmic rays, $\gamma$, and a correction
factor, $f_A$ which takes into account that some primaries and targets
are nuclei, and not protons.  The function ${\cal F}$ varies
slowly with energy, and is largely independent of the form of the
cosmic ray spectrum at ultrahigh energies, (see {\em e.g.}, \cite{Der86},
\cite{GS91}, \cite{Aha91}, \cite{BZ90} 
for details concerning diffuse emissivity
predictions.)  

Since the column density of hydrogen gas in molecular clouds is
well-established from radio observations, the uncertainty in
estimating the flux of diffuse gamma rays is dominated by the
uncertainty in the flux of hadronic cosmic rays within the cloud.  We
can reduce the effect of this uncertainty by considering the {\em
ratio} of gamma ray to cosmic ray fluxes. In the case of a uniform
flux of cosmic rays, this ratio will simply be proportional to the average
column density.  For example, using the value of ${\cal F}$ from
Aharonian (1991), the expected diffuse emission is given as:
\begin{equation} 
\frac{J_{\gamma}}{J_{\mbox{\tiny CR}}} \simeq 2 \times 10^{-5} \cdot
<N_{H}>_{22},
\end{equation} 
where $<N_{H}>_{22}\equiv <N_{H}>/(10^{22}$ cm$^{-2})$ is the average
column density of gas in any direction. In the plane of the outer
galaxy visible by CASA-MIA ($40^{\circ} \leq l
\leq 200^{\circ}$), $<N_{H}>_{22}$ varies between 0.8 and 1.2 for
$|b| \leq 5^{\circ}$ ({\em
e.g.}, \cite{Blo84}). In this case, the flux ratio,
$J_{\gamma}/J_{\mbox{\tiny CR}}$, depends only weakly on the exact
form of the cosmic ray spectrum, and is expected to be constant as a
function of energy to within a factor of 2 over the range from 100 to
400 TeV.  A realistic estimate of the emission based upon a
three-dimensional model of galactic column density
yields a similar result (\cite{Berz93}).  We note
that if the ambient flux of cosmic rays within a particular cloud is
stronger than the average flux measured at earth
({\em e.g.}, because the cloud is in close proximity to a source
of cosmic rays), then we would expect the diffuse gamma-ray flux to be
stronger than predicted.

\section{Observations and Analysis Method}

When ultrahigh energy cosmic rays and gamma rays strike the upper
atmosphere of the earth, they produce large cascades of
electromagnetic and hadronic particles.  For primary particle energies
around 100 TeV these particle cascades, called extensive air showers,
reach the ground and can be detected readily by surface scintillation
counters.  At ground level, the showers consist largely of electrons,
photons, and muons.  A vertically incident 100 TeV proton primary
produces a shower at ground level of roughly 30,000 electrons and 2000
muons.  The shower consists of a thin ``pancake'' of relativistic
particles, whose orientation preserves the direction of the incident
primary particle.  A sparse array of detectors sampling a small
fraction of the shower particles can thereby determine the total
number of particles in the shower and the arrival direction of the
primary particle with reasonable precision.  The combination of the
Chicago Air Shower Array (CASA) and the Michigan Muon Array (MIA)
comprise the largest and most sensitive such array built to date.

The CASA experiment consists of 1089 scintillation detectors laid out
on a square grid of 15m spacing.  The experiment is located at Dugway,
Utah (40.2$\arcdeg$ N, 112.8$\arcdeg$ W) at an atmospheric depth of
870 g/cm$^{2}$. The total collection area of CASA is roughly
230,000 m$^{2}$, of which 0.7\% is actually covered by scintillator.
The detectors have local analog and digital electronics to record the
number of particles and the shower arrival time at each
scintillator.  The number of particles detected can be used to
estimate the energy of the primary particle (cosmic ray or gamma ray).
By comparing the observed shower rate to the known cosmic ray flux, we
find that the median energy of gamma rays creating showers detectable
by CASA is approximately 120 TeV (\cite{casa_crab}).

At the Dugway site, beneath CASA, we have constructed a very large
array of muon counters -- the Michigan Muon Array (MIA).  This
array consists of 1024 scintillation counters buried three meters below
ground level.  The counters are grouped into sixteen ``patches'' and
have a total scintillator area of 2500 m$^{2}$.  The amount of earth
overburden above MIA results in a muon energy threshold of $\sim$ 0.8
GeV; electromagnetic punch-through is negligible for all but the very
largest showers.  A complete description of the 
CASA-MIA instrument can be found elsewhere (\cite{Bor94}).

In showers initiated by cosmic ray nuclei, muons are produced from
charged pion and kaon decays in the central hadronic core.  Showers
initiated by gamma rays do not produce muons in significant quantities
because the cross section for $\gamma$-air hadroproduction is much
smaller than the cross section for electron-positron pair production
($\gamma$-air $\rightarrow e^+e^-$).  In other words, in contrast to
cosmic ray nuclei, gamma rays usually interact electromagnetically,
and produce air showers with fewer charged mesons, and therefore fewer
muons at ground level.  On average, CASA-MIA detects approximately
nine muons for each cosmic ray air shower; in contrast, we expect to
detect only 0.25 muons for each gamma-ray shower, on average.

To search for diffuse concentrations of cosmic gamma rays, we employ
the technique developed by Matthews et al.~(1991).  The number
of muons resulting from a gamma-ray initiated shower should be, on
average, much less (by a factor of 30 to 40) than the number of muons
from a hadron-initiated shower of comparable energy.  We therefore
search for gamma-ray emission by looking for a localized excess of
cosmic rays which are muon-poor with respect to the number of muons
expected from typical hadron-initiated showers (\cite{Cov91}). 

We parameterize the muon content of each shower by the
quantity:
\begin{equation}
r_{\mu}\equiv
\log_{10}\frac{(n_{\mu\mbox{\small\_obs}})}{(n_{\mu\mbox{\small\_exp}})},
\end{equation}
where $n_{\mu\mbox{\small\_obs}}$ and $n_{\mu\mbox{\small\_exp}}$
represent the numbers of muons actually observed and the number of
muons {\em expected}\/ for hadronic showers.  We determine the
expected number of muons on a
shower-by-shower basis by parameterizing the lateral density of muons
in terms of the function 
$\rho_{\mu}({R},N_{e},\theta_{z})$
described by \cite{Gri60},
where ${R}$ is the distance to the
shower core, $N_{e}$ is the electron shower size, and $\theta_{z}$ is
the zenith angle of the incident shower.

\begin{figure}
\plotone{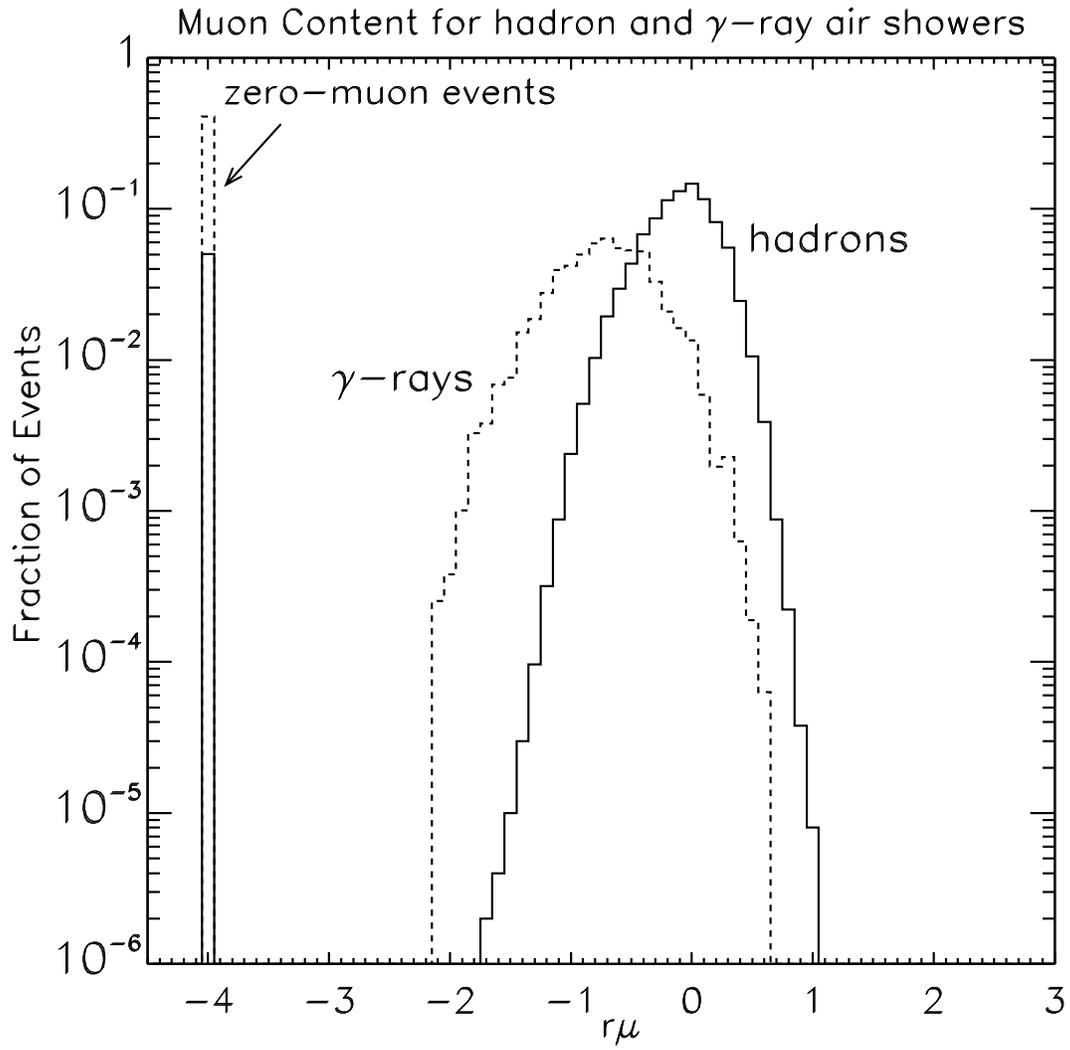}
\caption{The $r_{\mu}$ distribution of simulated gamma ray (dashed)
vs. detected hadronic (solid) cosmic rays.}
\label{f:compare}
\end{figure}

Having parameterized the muon content in this manner, we consider the
expected distribution of $r_{\mu}$ for both hadronic and gamma-ray
primaries.  For hadrons, we expect that the distribution of $r_{\mu}$
is characterized by the actual array data, since the vast majority of
air showers (at least 99.9\%) are known to be hadronic.  For
gamma rays we must invoke Monte Carlo simulations to determine the
expected distribution of $r_{\mu}$. The distribution for gamma rays
will be shifted significantly towards lower values of $r_{\mu}$ and
will include a much larger fraction of events with zero muons
detected.  The $r_{\mu}$ distributions for both hadronic and simulated
gamma-ray air showers are shown in Figure~\ref{f:compare}.  

We define as {\em muon-poor} those events which have a value of
$r_{\mu}$ less than a value chosen to optimize sensitivity to gamma
rays.  This sensitivity scales as $h_{\gamma}/\sqrt{h_{\mbox{\tiny
CR}}}$, where $h_{\gamma}$ and $h_{\mbox{\tiny CR}}$ represent the
fraction of gamma rays and cosmic rays retained in the muon-poor sample,
respectively.  The optimal value of $r_{\mu}$ depends weakly on the
source declination and the energy of the selected events
(\cite{casa_binary}). 

To determine if there is evidence for gamma-ray emission from a
localized candidate source region of the sky, we examine the
distribution of $r_{\mu}$ for showers within this region, and we
search for an excess of muon-poor events. The excess is determined
relative to a comparison region which presumably contains a
background of uniformly hadronic showers. To compare the $r_{\mu}$
distributions from source and background regions, we normalize so as
to match the total number of events with $r_{\mu} > 0.0$, assuming that
all such events are purely hadronic.  We note that this technique
has no sensitivity whatsoever to the presence of any isotropic
flux of gamma rays which may be present in our data. 

To the extent that our parameterization of the muon lateral
distribution applies generally to all showers, we expect that the
distribution of $r_{\mu}$ should remain identical for all hadronic
showers at a fixed overburden throughout the atmosphere.  However,
since the overburden changes rapidly with zenith angle to the shower,
we expect that the shape of the $r_{\mu}$ distribution will depend
upon the zenith angle.  If uncompensated, these systematic changes in
the shape of the $r_{\mu}$ distribution can have a significant impact
upon our estimate of muon-poor showers.

We have developed a technique for characterizing and
removing the systematics in the $r_{\mu}$ distribution.  We divide the
entire data sample into a series of coarse bins in time (about one bin
per day) and horizon coordinates (8--10 bins each in elevation and
azimuth).  For each source event, we generate comparison background
events by sampling a value of $r_{\mu}$ from the distribution
corresponding to the appropriate horizon coordinate bin.  Thus,
systematic changes to the shape of the $r_{\mu}$ distribution that are
local to the events in each sky bin will appear in both source and
background distributions. This technique removes most systematic
effects, provided that the bin size is chosen so that the $r_{\mu}$
distribution obtained from the data represents a good approximation of
the $r_{\mu}$ distribution for the entire bin (\cite{Cov94}).

\begin{figure}
\plotone{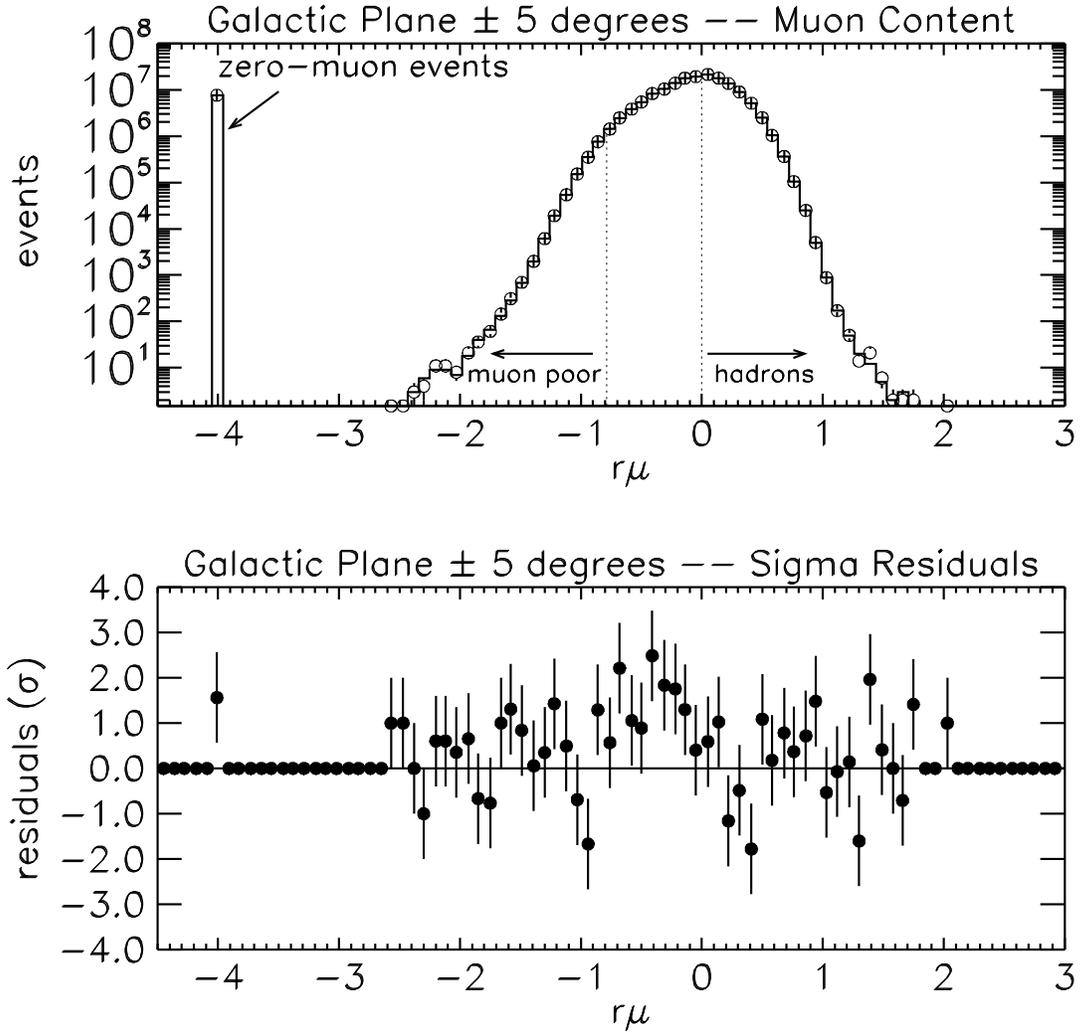}
\caption{The upper plot shows the muon content ($r_{\mu}$
distribution) for CASA-MIA air shower events arriving from the direction
of the galactic plane ($|b_{II}| < \pm 5\arcdeg$). Points with error bars
represent the on-source distribution, while histogram lines represent
the off-source distribution.  The lower plot indicates the on-source minus
off-source residuals (sigmas per bin).  Evidence for the presence of
ultra-high energy gamma rays would be indicated by an excess of
muon-poor events.  A statistically insignificant excess ($1.63 \sigma$)
is seen.}
\label{f:rmu_plot}
\end{figure}

Figure~\ref{f:rmu_plot} graphically demonstrates the application of
this technique to real data.  The $r_{\mu}$ distribution for candidate
diffuse source events within the galactic plane ($\pm$ 5$\arcdeg$) is
plotted against the $r_{\mu}$ distribution for the off-source
background region.  Analysis of residuals shows no major differences
between the source and background distributions, and gives us
confidence that the systematics have largely been removed.  We
attribute the excess in the chi-squared ($\chi^{2}_{\nu}$ =1.25, $\nu$
= 51 d.o.f) to either an unremoved systematic effect, or the marginal
presence of a possible signal.  If there were a detectable flux of
gamma rays from this region of the galactic plane,
Figure~\ref{f:rmu_plot} would have shown an excess of signal events
over background on the left-hand side of the plot.  Indeed, there is
an excess (1.63 standard deviations in the plot shown) which, while
consistent with the expected flux, is not significant enough to claim
as a detection.

\section{Results}

\begin{figure}
\plotone{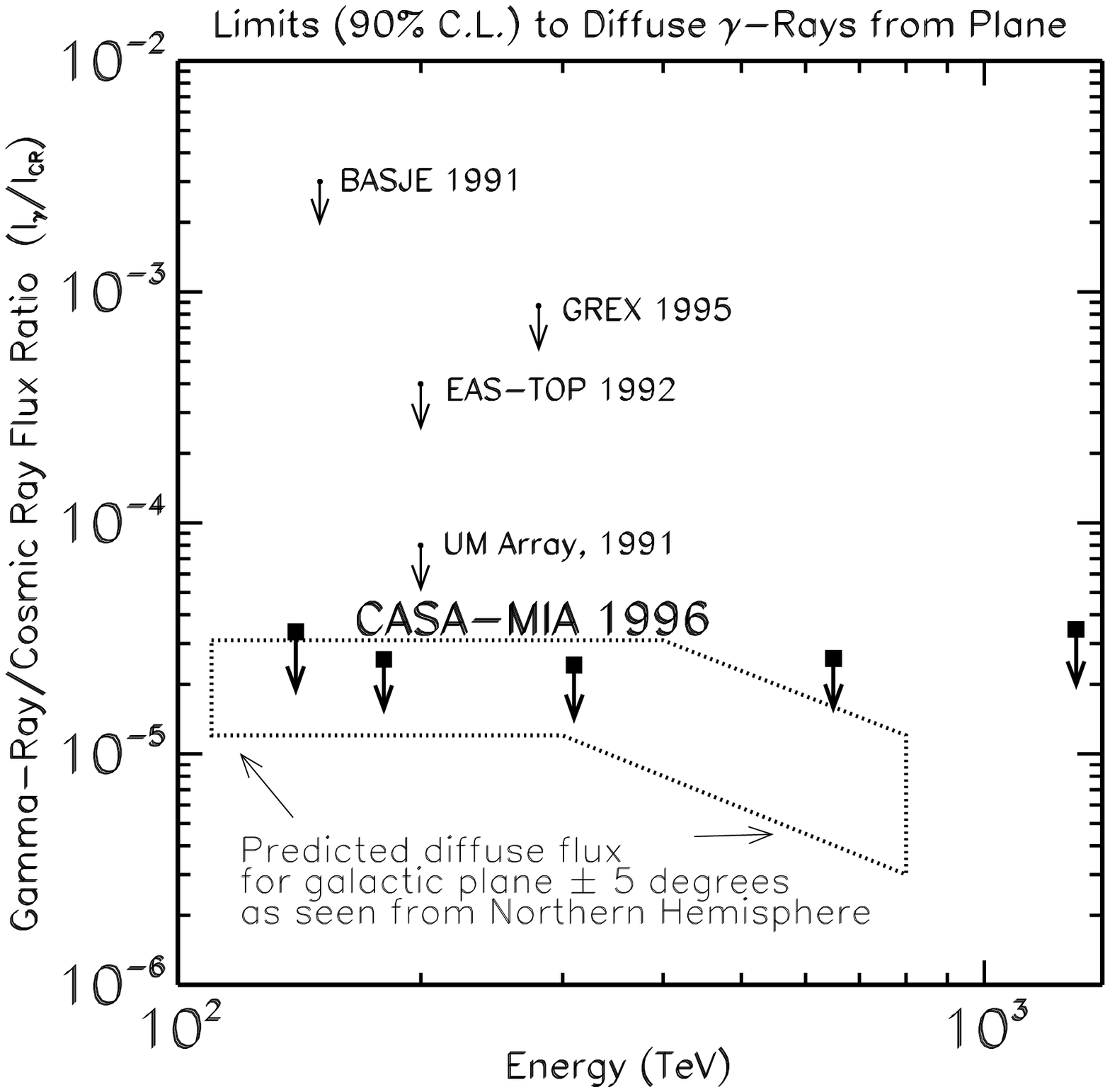}
\figcaption{CASA-MIA sensitivity
to diffuse gamma ray emission
from the central plane of the galaxy ($|b_{II}| <\pm 5\arcdeg$, $50\arcdeg <
\ell_{II} < 200\arcdeg$).  Sensitivities are given in terms of
the fraction of gamma rays relative to the detected all-particle flux
of cosmic rays at the earth. Also shown are limits from previous
experiments (BASJE--\protect\cite{BASJE},
EAS-TOP--\protect\cite{Agl92},
UM--\protect\cite{Mat91}.) Predicted flux from \protect\cite{Aha91}.}
\label{f:diffratio}
\end{figure}

We apply this analysis technique to search for evidence of gamma rays
from the galactic plane region in the energy range from 140 to 1300
TeV.  No significant evidence for diffuse emission is found.  Upper
limits to diffuse flux from the plane
of the galaxy are calculated taking into account the slight loss in
sensitivity ($< 10\%$) due to the finite angular resolution of CASA.
We present 90\% confidence level upper limits on the flux for
emission for a range of emission disk thicknesses ($|b_{II}| <
2\arcdeg$, $5\arcdeg$, and $10\arcdeg$) for comparison to previous
predictions and experiments.  (For example, \cite{Agl92} have claimed
that the band $|b_{II}| < 2\arcdeg$ best matches the expected UHE
emission, based on COS-B results, while we believe that the EGRET
results demonstrate that $|b_{II}| < 5\arcdeg$ is the most appropriate
search region).  The CASA-MIA flux limits are plotted in
Figure~\ref{f:diffratio}.  Also shown are upper limits from previous
experiments and a theoretical prediction for the expected flux.  We
note that cosmic ray rejection improves rapidly with energy, thus
compensating for reduced statistics, so that the sensitivity of
CASA-MIA to diffuse emission is relatively constant in the region from
200 to 1000 TeV.

\begin{deluxetable}{cccc}

\tablecaption{Limits to Diffuse Emission}
\tablewidth{0pt}
\tablecolumns{4}
\tablehead{
\colhead{Region $(50\arcdeg <\ell_{\mbox{\tiny II}} < 200\arcdeg$)}  &  
\colhead{Median Energy (TeV)}  &
\colhead{Significance ($\sigma$)} &
\colhead{$J_{\gamma}/J_{CR}$ 90\% C.L. ($10^{-5}$)}
}
\startdata
$ -2\arcdeg < b_{\mbox{\tiny II}} <  2\arcdeg$  & 
                                   140          & +1.78 & 7.2  \nl
                                &  180          & +1.81 & 3.8  \nl
                                &  310          & +2.56 & 5.2 \nl
                                &  650          & +1.12 & 3.2 \nl
                                &  1300         & +0.07 & 4.6 \nl   
        &          &         &          \nl

$ -5\arcdeg < b_{\mbox{\tiny II}} <  5\arcdeg$  & 
                                   140          & +1.63 & 3.4  \nl
                                &  180          & +0.08 & 2.6  \nl
                                &  310          & +0.86 & 2.4 \nl
                                &  650          & +1.60 & 2.6 \nl
                                &  1300         & +0.06 & 3.5 \nl   
        &          &         &          \nl
$ -10\arcdeg < b_{\mbox{\tiny II}} <  10\arcdeg$ &
                                   140          & +2.39 & 2.8 \nl
                                &  180          & +1.79 & 2.2 \nl
                                &  310          & +0.87 & 2.3 \nl
                                &  650          & +0.91 & 1.8 \nl
                                &  1300         & -0.56 & 2.3 \nl   
\enddata
\tablecomments{Tabulated upper limits to diffuse gamma-ray emission from the
plane of the galaxy. Although positive excesses are seen, we do not
view these as statistically significant enough to claim detections.
Flux limits are tabulated for bands along the galactic plane from
$|b_{\mbox{\tiny II}}| < 2\arcdeg$ to $|b_{\mbox{\tiny II}}| <
10\arcdeg$.  Median energy is quoted for integral flux limits.
Selected spatial regions and energy bands are not statistically
independent.}
\label{t:limits}
\end{deluxetable}

\section{Discussion}

The observed weak excess of muon-poor events in several of the
selected spatial and energy ranges could be taken as evidence for the
presence of diffuse emission from the galactic plane.  Indeed, if the
excess is interpreted as a positive gamma-ray signal, the inferred
flux is consistent with current model predictions.  However,
in view of the uncertainty in assessing
the impact of any remaining systematics, we interpret
the observed excesses as not statistically significant enough to
warrant a claim for detection, and we use our results to place upper
limits on the predicted emission.  We note that future prospects for
increased sensitivity to verify or refute the presence of a possible
signal at this level are not good, since this would require a much
larger experiment and/or a much longer running time, neither of
which is foreseen.

The CASA-MIA flux limits place strong constraints on models for
diffuse emission.  These limits are below the predictions of some
earlier models ({\em e.g.}, \cite{BZ90}) and are approximately a
factor of two above the minimum emission predicted by more recent
models ({\em e.g.}, \cite{Aha91}, \cite{Berz93}).  These predictions
are based upon diffuse emission due to the process of normal cosmic
ray nucleon-nucleon interactions on passive gas targets only.
Therefore we conclude that no other emission process is likely to
dominate at ultrahigh energies.  Some models for cosmic ray
acceleration suggest that gamma rays from clouds in proximity to SNR
may be enhanced and/or spectrally hardened ({\em e.g.}, \cite{DAV}).
Others have suggested that inverse Compton scattering may be a
significant component of galactic diffuse emission at ultrahigh
energies ({\em e.g.}, \cite{AA95}).  A report by \cite{Blo91} based
upon an analysis of the COS-B data suggests evidence for a spectral
hardening at energies above 1 GeV ($\Delta_{\gamma} =0.4$) for the
outer galaxy; however this result has not been confirmed by EGRET
(\cite{Hun97}).  The CASA-MIA upper limits rule out any such spectral
hardening above 200 TeV in the outer galaxy with $\Delta_{\gamma} >
0.1$.

\acknowledgments

{\small The authors acknowledge the help and support of the members of the
University of Utah's Fly's Eye Collaboration and the staff of the
Dugway Proving Ground. We thank M. Cassidy for technical support at
Dugway.  We also wish to thank L. Nelson, M. Oonk, P. Burke,
S. Golwala, M. Galli, J. He, P. Rauske, Z. Wells, H. Kim,
M. Pritchard, and K. Riley for help in processing the data.  We thank
F.A. Aharonian for helpful discussions. We thank S.D. Hunter for
helpful discussions related to diffuse emission from EGRET. This work
is supported in part by the National Science Foundation and the
U.S. Department of Energy. JWC and RAO acknowledge the support of the
W.W. Grainger Foundation.}

\end{document}